\documentclass[twocolumn,prl,amsmath,aps,groupedaddress,superscriptaddress,
floatfix,showpacs]{revtex4}

\usepackage{epsfig}
\usepackage{graphics}
\usepackage{amsmath}
\usepackage{verbatim}

\newcommand{\bra}[1]{\langle #1|}
\newcommand{\ket}[1]{|#1\rangle}

\begin{document}
\title{Heisenberg limited measurements with superconducting circuits}
\author{Alexandre Guillaume}
\email{Alexandre.Guillaume@jpl.nasa.gov}
\affiliation{Jet Propulsion Laboratory, California Institute of Technology,
Mail Stop 302-306, 4800 Oak Grove Drive, Pasadena, CA 91109-8099}
\author{Jonathan P. Dowling}
\affiliation{Hearne Institute for Theoretical Physics, Department of
Physics and Astronomy, Louisiana State University,202 Nicholson Hall,
Tower Drive, Baton Rouge, LA 70803-4001}
\affiliation{Inst.~for Quantum Studies \& Dept. of Physics, Texas
A\&M University, College Station, Texas 77843-4242}
\date{\today}
\begin{abstract}
We describe an assembly of \textit{N} superconducting qubits
contained in a single-mode cavity. In the dispersive regime, the
correlation between the cavity field and each qubit results
in an effective interaction between qubits that can be used to
dynamically generate maximally entangled states. With only collective
manipulations, we show how to create maximally entangled quantum
states and how to use these states to reach the Heisenberg limit in
the determination of the qubit bias control parameter (gate charge for
charge qubits, external magnetic flux for rf-SQUIDs).
\end{abstract}
\maketitle

The description of the interaction between atoms and quantized modes of the
electromagnetic field in a cavity is called cavity quantum electrodynamics
(cQED). The first experimental studies used flying Rydberg atoms in a rf
resonator \cite{raimond2001}. With the advent of quantum computing several other
implementations were developed to mimic the quantum properties of atoms. Among
those, solid-state implementations are especially interesting since they offer
several advantages over real atoms: these artificial atom properties can be
tailored and their number and location are fixed. We concentrate our discussion
on superconducting circuits. Superconducting implementations of quantum
computing attracted a lot of attention in recent years because they are
inherently scalable and single qubit operations have been demonstrated with
classical coherent control in a variety of qubit types. Several theoretical
studies have treated the interaction of superconducting qubit(s) with a quantized
electromagnetic field.
The proposal of an on-chip cQED experiment using a Cooper-pair box as the
artificial atom strongly coupled to a one-dimensional cavity \cite{blais2004}
is especially interesting since it was followed by several experiments.
First, the strength of the coupling was shown to be indeed stronger than the
different decay constants so that the vacuum Rabi splitting was observed
\cite{wallraff2004}. Subsequently the ac-Stark shift  was measured using a
quantum non-demolition technique (QND) \cite{schuster2004} and the decoherence
time $T_2$ evaluated from Ramsey fringe experiments \cite{wallraff2005}.

In a cavity with a high quality factor, the photons can serve as an information
bus between several qubits and therefore create correlations between distant
qubits. Beside its fundamental interest and applications to quantum information
processing, entanglement offers the additional advantage of allowing improved
sensitivity in a quantum-limited measurement. Compared to a measurement made
with only one sensor, the sensitivity is improved by $1/\sqrt{N}$ when $N$
classical sensors are used to measure the same quantity. Now, if $N$ quantum
sensors are \textit{coherently} coupled, the sensitivity improvement can reach
$1/N$. This ultimate sensitivity is referred to as the Heisenberg limit.
Techniques to create and control multiparticle entangled states necessary to
reach this limit are already available in different implementations.
For instance,
interferences between two different polarizations (modes) of three and four
photon entangled states have been observed \cite{mitchell2004,walther2004} and
an experiment performed on an assembly of three beryllium ions demonstrated a
spectroscopic sensitivity improvement \cite{leibfried2004}.
Today's solid-state sensors do not take advantage of the improvement made
possible by quantum correlations.
Using entanglement in a superconducting implementation to enable such Heisenberg
limited measurements, could therefore revolutionize sensor technology with, for
instance, electrometers and magnetometers.

In this work, we use techniques from other fields (atomic QED, ion traps,
quantum optics) to describe the interactions and manipulations needed in a 
superconducting circuit to perform a Heisenberg limited measurement.
We propose to use the photon-qubit interaction to create an effective
interaction between distant superconducting qubits. This interaction is used to
generate maximally entangled states which in turn are used to beat the standard
quantum limit when measuring the natural frequency of the system. We show how
this results in a Heisenberg limited estimation of the qubit bias control
parameter.

Most superconducting qubits can be described by the following single qubit
Hamiltonian \cite{makhlin2001a}:
\begin{equation}
	\label{eq:sctls}
	H_Q=-\frac{B_z}{2}\,\sigma_z-\frac{B_x}{2}\,\sigma_x,
\end{equation}
with the bias $B_z$ depending linearly on the dimensionless control parameter
$\lambda$: $B_z=b_z(1/2-\lambda)$. This Hamiltonian is an approximation valid
around a symmetry point, obtained for $\lambda=1/2$, called the degeneracy
point. The parameters $B_z$, $B_x$, $b_z$ and $\lambda$ for the two main types
of superconducting qubits, can be found in the review article \cite{makhlin2001a}
on Josephson based devices. The bias is controlled with an electric gate charge
$n_g$ in a Cooper-pair box (CPB) or with an external magnetic flux $\phi_x$ in
a rf-SQUID. 

We now consider that the qubit is contained in a cavity described by the
Hamiltonian $H_c=\omega_c (a^{\dagger}a+1/2)$. The quantized cavity mode adds
an incremental contribution $\delta\lambda$ to the bias so that
$\lambda+\delta\lambda=\lambda+\lambda_c(a^{\dagger}+a)$. The single qubit
Hamiltonian now reads:
\begin{equation}
	\label{eq:tls+cav}
	H_{Q-C}=-\frac{B_z}{2}\sigma_z-\frac{B_x}{2}\sigma_x
	+\frac{b_z \lambda_c}{2}(a^{\dagger}+a)\sigma_z.
\end{equation}
We assume the strong coupling limit $g\gg\kappa,\gamma$. We neglect the cavity
decay $\kappa$ and excited state decay $\gamma$ for the moment. Experiments
on a single CPB in a cavity \cite{wallraff2004} support this assumption
($g\approx10\, \kappa$).
We rewrite this Hamiltonian in the eigenbasis of (\ref{eq:sctls}) which is done
by performing a rotation of $\theta$ about the $y$ axis. The mixing angle
$\theta$ is given by $\tan\theta=B_x/B_z$. The resulting individual contribution
is then summed $N$ times to describe an assembly of $N$ identical qubits in a
single mode cavity:
\begin{gather}
	H_N=\Omega \, S_z+\omega_c (a^{\dagger}a+1/2)
	\nonumber \\
	\label{eq:Ntls+cav}
	+2g (a^{\dagger}+a)(S_z\cos \theta+S_x\sin \theta),
\end{gather}
where $2g=-b_z \lambda_c$, $\Omega=\sqrt{B_x^2+B_z^2}$ and
$2\vec S=\sum_{i=1}^N \vec{\sigma_i}$.
Of course, the assumption of exact parameter degeneracy is rather
stringent. Our intent here is to study the ideal case to obtain simple
analytic expressions. 
When $\theta=\pi/2$ and the rotating wave approximation is made, equation
(\ref{eq:Ntls+cav}) takes the form of the $N$-particle Jaynes-Cummings
Hamiltonian. When $\theta=0$, equation (\ref{eq:Ntls+cav}) describes a harmonic
oscillator with a conditional displacement, \textit{e.g.} a displacement that
depends on the total state of the system $S_z$. However, the limit
$\theta\rightarrow 0$ is reached when the  bias $B_z$ is maximum and
$B_x\rightarrow 0$. In this case, the two-level approximation (\ref{eq:sctls}),
and hence equation (\ref{eq:Ntls+cav}), is not justified. We assume that the
qubit-cavity detuning $\Delta=\Omega-\omega_c$ is much larger than the qubit
coupling $g$ to the cavity. We write $S_x=(S_+ +S_-)/2$ and neglect the rapidly
oscillating terms.
Hamiltonian (\ref{eq:Ntls+cav}) can be approximately diagonalized with a
polaronic transformation
$U=\exp\big( \frac{g\sin\theta}{\Delta} (a S_{+}-a^{\dagger}S_{-}) \big)$:
\begin{gather}
	\tilde H_N=U H_N U^{\dagger}\approx
	\Omega \, S_z+\omega_c (a^{\dagger}a+1/2)
	\nonumber \\
	\label{eq:HNdiag}
	+\chi	(S^2-S_z^2+S_z+2a^{\dagger}a S_z).
\end{gather}
We define $\chi=(g\sin \theta)^2/\Delta$.
A more complete diagonalization would require to also take into account
the displaced harmonic oscillator with $U_d=\exp(\frac{2g\cos\theta}{\omega_c}
(a-a^{\dagger})S_{z})$ and perform the transformation 
$U_d U H_N U^{\dagger} U_d^{\dagger}$. However, this complete transformation
introduces terms proportional to $g^2/\omega_c$ which we neglect because we
are interested in the regime where $g\ll\Delta\ll\omega_c$. We also neglect
terms in $\chi^2$ that would appear with the complete transformation. Hence,
equation (\ref{eq:HNdiag}) is valid to first order in $\chi$.
The coupling of the qubits with the resonator induces a shift in the qubit
frequency and a state dependent shift in the resonator frequency. These Lamb
and ac-Stark shifts (respectively) were predicted in the case of a single CPB
in a cavity \cite{blais2004}.

The novelty of equation (\ref{eq:HNdiag}) compared to previously published
results is the appearance of an effective interaction $H_{sz}=\chi S_z^2$
between $N$ qubits mediated by a cavity photon.
This interaction is also known as a \emph{one-axis twisting} interaction
\cite{ueda1993} because it twists around the $z$ axis the quantum fluctuations
of the total spin $\vec S$ (for a system with several spins). Because of this
feature, the interaction $H_{sz}$ is a key element in non-optical
implementations of Heisenberg limited estimations. It has been used extensively
in ion traps experiments to generate maximally entangled states
\cite{sacket2001,kielpinski2001,sacket2000}. It has been shown to be possible
to utilize these states to perform an optimal frequency measurement
\cite{bollinger1996} and improve the estimation of rotation angles
\cite{meyer2001}. More recently, a method involving only collective
manipulations has been used to perform precision spectroscopy on an assembly of
six beryllium ions \cite{leibfried2005}. 
The first step consists in generating the maximally entangled state
$\ket{\psi_m}$ (see for instance reference \cite{molmer1999a}) using the time
evolution of $H_{sz}$ over a time $t_{sz}=\pi/2\chi$:
\begin{gather}
	\ket{\psi_m}=e^{-i \frac{\pi}{2} S_z^2}\ket{-N/2}_x=
	\nonumber \\
	\label{eq:GHZstate}
	\frac{1}{\sqrt{2}}\Big(\ket{-N/2}_x+i^{N+E}\ket{+N/2}_x\Big).
\end{gather}
When the number of qubit $N$ is odd, another rotation $e^{i \frac{\pi}{2}S_z}$
is needed in addition to the $e^{i \frac{\pi}{2}S_z^2}$ \cite{molmer1999a}.
We take the parity into account in equation (\ref{eq:GHZstate}) by 
setting the quantity $E$ to $2$ ($1$) when $N$ is odd (even resp.).
Initially all the qubits are in their ground state, so that the total wave
function describing all the qubits is $\ket{J=N/2,M=-N/2}_z\equiv
\ket{-N/2}_z=\ket{\downarrow}_1\ket{\downarrow}_2\cdots \ket{\downarrow}_N$
(we do not consider the field part of the wave function at this point).
In order to prepare a quantum state according to equation (\ref{eq:GHZstate}),
we need to rotate the initial state $\ket{-N/2}_z$ around the $x$ axis.
Therefore, we define the operator 
$U_N=e^{i \frac{\pi}{2} S_x}e^{-i H_{sz} t_{sz} }e^{-i \frac{\pi}{2} S_x}$
for notation convenience.
The average of the total spin vector calculated with the wave function
$\ket{\psi_m}$ is zero, $\langle \vec S \rangle=0$. Thus, the natural
choice of $S_z$ ($\vec S$) as the observable to be measured can not be made. 
Bollinger \textit{et al.} showed that the parity operator
$\prod_{i=1}^N \sigma_{z_i}$
was an adequate observable that could be measured with the state
(\ref{eq:GHZstate}). However the measurement of this operator for a large
number $N$ of qubits is difficult since it requires distinguishing odd
and even numbers of particles in state $\ket{\downarrow}$. A method involving
only collective manipulations has been proposed in \cite{leibfried2004} to
circumvent this problem. First the maximally entangled state is constructed.
In our case the sequence $U_N$ is applied to the initial state $\ket{-N/2}_z$.
Afterward, the system evolves freely during a period $T$, obeying the
dynamics defined by the Hamiltonian $\tilde H_N$. Finally, another application
of the sequence $U_N$ transfers the phase information, $N\phi/2$, into an
amplitude information of either state $\ket{+N/2}_z$ or $\ket{-N/2}_z$:
\begin{gather}
		\ket{\psi}=U_N e^{-i \tilde H_N T}U_N \ket{-N/2}_z=
		\nonumber \\
		-i \sin\Big( \frac{N}{2}\, \phi \Big)\ket{-N/2}_z+
		\nonumber \\
		\label{eq:seq}
		i^{N+E} \cos\Big( \frac{N}{2}\, \phi \Big)\ket{+N/2}_z.
\end{gather}
A measurement will collapse the wave function $\ket{\psi}$ on either
state $\ket{+ N/2}_z$ or state $\ket{- N/2}_z$ with probability
$P_{\uparrow}=\frac{1}{2}(1+\cos(N \phi))$ or
$P_{\downarrow}=\frac{1}{2}(1-\cos(N \phi))$.

We propose to use cavity spectroscopy to infer the state of the qubits.
Assuming that there is a finite but small cavity decay rate $\kappa$, a signal
at frequency $\omega_c$ sent in the cavity will experience a phase shift when
it is transmitted. Solving the Heisenberg equation for the field creation
operator, this phase shift $\vartheta$ is given by
$\tan \vartheta=\pm(2\chi N)/\kappa$. The probability $P_{\uparrow}$ is
extracted from the time dependence of $\vartheta$.
A measurement scheme as been proposed based on this principle to perform a
quantum non-demolition measurement of the state of a single Cooper-pair box
contained in a cavity \cite{blais2004}. The difference is that because the
coupling to the cavity is $\sqrt N$ stronger, the phase shift $\vartheta$ is
larger than in the single qubit case.

The main motivation to use $N$-particle maximally entangled state to perform
a spectroscopy measurement is to be able to relate the $N$ fold frequency 
increase to the phase uncertainty. The uncertainty on a parameter $\zeta$ can
be estimated from the error propagation formula
$\delta\zeta=\Delta \hat A/ |\partial\langle\hat A\rangle/\partial\zeta|$
by measuring the operator $\hat A$.
We introduce the projection operator 
$\hat A=\ket{+N/2}\bra{+N/2}$ so that the quantity we propose to measure
$P_{\uparrow}$ is the average of $\hat A$ over the state $\ket{\psi}$
of equation (\ref{eq:seq}). The variance $\Delta \hat A^2$ is then simply 
given by $P_{\uparrow}(1-P_{\uparrow})$ (second moment of the Bernoulli
distribution) which is equal to $(\sin(N \phi)/2)^2$. The denominator of
the error propagation formula, $|\partial P_{\uparrow}/\partial\phi|$, is
$N\times\sin(N \phi)/2$. Hence, the measurement of $P_{\uparrow}$ leads to an
estimation of the phase uncertainty $\delta \phi$ equal to $1/N$.
The phase acquired during the free evolution, by the qubit part of the wave
function, is $(\Omega+\chi+2\chi \bar{n})\times T$ where $\bar{n}$ is the
average photon number in the cavity. The contribution of the frequency shifts is
negligible in the expression of the phase uncertainty. Thus, the frequency
uncertainty is given by $\delta\Omega=1/(N T)$.
In superconducting circuits, the parameter $\lambda$ controls the level spacing
$\Omega$ and therefore the uncertainties of both quantities can be related
through the  following relation:
\begin{equation}
	\delta\Omega=
	\frac{b_z |B_z|}{\sqrt{B_x^2+B_z^2}}\, \delta\lambda=
	b_z |\cos\theta| \, \delta\lambda
\end{equation}

Hence, a measurement of the frequency $\Omega$ performed with the sequence
of operations defined by equation (\ref{eq:seq}) results in a Heisenberg limited
measurement of the parameter $\lambda$, \textit{e.g.} in an improvement of the
uncertainty $\delta\lambda=1/(N T\times b_z |\cos\theta|)$ associated with the
estimation of $\lambda$. This method can be used to improve the estimation of
the gate charge $n_g$ (or the external bias flux $\phi_x$) in a system composed
of $N$ Cooper-pair boxes (or rf-SQUIDs resp.) that are coherently coupled. The
energies $b_z$ and $B_x$ can be determined from a preliminary spectroscopy
experiment. Sweeping $\lambda$ when applying a periodic signal will flip the
qubits when the frequency is resonant.

Our scheme is useful \emph{away} from the degeneracy point as at this point
$\cos\theta=0$. However, one should bear in mind that the coupling
$\chi$ decreases as the operating point is moved away from the degeneracy point,
so a trade-off should be made to operate away, while not too far from this point.
The limited validity range around the degeneracy point is not a limitation in a
system composed of $N$ Cooper-pair boxes since the different coherence times
decrease with the distance from this point \cite{ithier2005} and therefore the
operation far away from it, is not adequate to observe coherent effects.

To aid the understanding of our scheme in particular and of Heisenberg limited
measurements in general, we wish to emphasize the difference between the quantum
observable measured and the parameter that the method allows to determine with a
better precision. The quantum observable defines the type of superconducting
sensors or qubit. To simplify the discussion, let's say there are mainly two
types of superconducting devices, electric charge and magnetic flux sensors.
The quantum observable measured is then either the electric charge $\hat n$ or
the magnetic flux $\hat \phi$. 
Now, the Hamiltonian of the system can be tuned or controled
with a classical (continuous) variable which is the gate charge $n_g$ in a charge
qubit or the external magnetic flux $\phi_x$ in a flux qubit. A Heisenberg
limited measurement consists of evaluating the uncertainty of the 
parameter ($\delta n_g$ or $\delta \phi_x$) for a given value of this parameter
with a measurement of the quantum observable ($\hat n$ or $\hat \phi$).
The parameters usually estimated in a Heisenberg limited measurement are either
an energy splitting, a rotation angle or a phase delay. Therefore, by
establishing for the first time a relation between other parameters, such as the
electric charge or the magnetic flux, and a quantum measurement, we show how a
Heisenberg limited measurement can have some applications in sensor technology.

In this work, we described a collection of $N$ superconducting qubits contained
in a single mode cavity. Besides the usual shifts in the qubit and resonator
frequencies, we find that the effective Hamiltonian contains a term $\chi S_z^2$
describing the interaction between all the qubits. This interaction can be used
to dynamically prepare maximally entangled states. We adapt a method used in ion
traps to demonstrate the use of these states to reach the Heisenberg limit in
the determination of the qubit frequency. Finally, we show that the parameter
that controls the energy spacing can be estimated with an uncertainty that
scales inversely with the number $N$ of qubits. Hence our work establishes the
first formal relation between, either the electric charge or the magnetic flux
in an assembly of superconducting devices, and the so-called Heisenberg limited
fluctuations.

\begin{acknowledgments}
The research described in this paper was carried out at the Jet Propulsion
Laboratory, California Institute of Technology, under a contract with the
National Aeronautics and Space Administration.
A.G. would like to acknowledge the National Research Council and NASA code S.
Also J.P.D. would like to acknowledge the Hearne Foundation, the National
Security Agency, the Disruptive Technologies Office and the Army Research
Office.
\end{acknowledgments}


\end{document}